\documentclass[aps,prb,longbibliography,reprint]{revtex4-2}

\usepackage[english]{babel}
\usepackage[utf8]{inputenc}
\usepackage[colorinlistoftodos, color=green!40, prependcaption]{todonotes}
\usepackage{graphicx}
\usepackage{bm}

\usepackage{amsmath}

\usepackage{xcolor}

\begin{document}

\title{Numerical Simulation of Radiative Transfer of Electromagnetic Angular Momentum  }

\author{B.A. van Tiggelen and R. Le Fournis}
\email[]{Bart.Van-Tiggelen@lpmmc.cnrs.fr}
\affiliation{Univ. Grenoble Alpes, CNRS, LPMMC, 38000 Grenoble, France}

\date{\today} 

\begin{abstract}
We present numerical simulations of light emitted by a source and scattered by surrounding electric dipoles with Zeeman splitting. We calculate the leakage of electromagnetic angular momentum to infinity.
\end{abstract}


\maketitle

\section{Introduction}

Optical sources radiate electromagnetic energy with a rate that depends on the  local density of radiative states (LDOS) near the source and at the emitted frequency of the source \cite{ldos}. This statement is  true classically and recalls Fermi's Golden Rule in quantum mechanics \cite{ldos2}. The LDOS is affected by the environment, and can be either dielectric, structured, gapped in frequency, or disordered. If the environment is magneto-active, induced by the presence of an externally applied magnetic field $\mathbf{B}_0$, the source can also radiate angular momentum (AM) with direction $\mathbf{B}_0$ into space.

\begin{figure}
\includegraphics[width=5cm, angle=0]{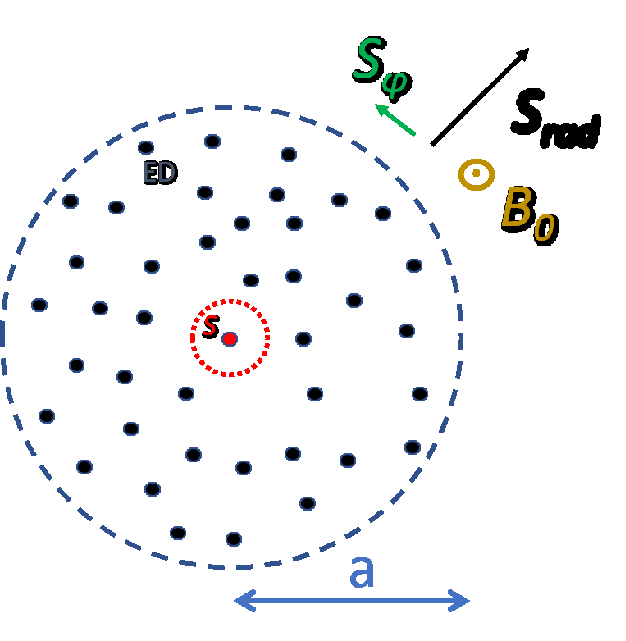}
\caption{The geometry considered in this work. A source emits light into a disordered environment containing $N$ electric dipole scatterers. In the presence of a magnetic field $\mathbf{B}_0$ a magneto-transverse component $\mathbf{S}_\phi$ of the Poynting vector appears outside the medium that carries electromagnetic angular momentum. This angular momentum propagates to infinity and a torque is exerted on scatterers and source.}
\label{setup}
\end{figure}

The first study \cite{pinwheel} of this phenomenon used the phenomenological concept of radiative transfer and in particular the role of the radiative boundary layer of an optically thick medium to argue that the Poynting vector has two components in the far field. The first is the usual energy flux,  purely radial, and decays with distance $r$ from the object as $1/r^2$. The second is magneto-transverse and circulates energy around the object (see Fiure~\ref{setup}). This component decays faster as $1/r^3$  but has finite angular momentum constant with distance that  travels away from the object. Our second study \cite{OL} demonstrated that this leakage is not restricted to multiple scattering and also exists when a homogeneous magneto-birefringent environment surrounds the source. In this case 1) the radiation of AM results in a torque on the source and not on the environment; 2)  it depends sensitively on the nature of the source with huge differences between e.g. an electric dipole source and a magnetic dipole source, and 3) geometric ``Mie" resonances can enhance the effect much like the Purcell effect does in nano-antennas \cite{purcell}.

Our latest study \cite{PRBAM} considered numerically a spherical environment filled with small resonant electric dipole scatterers. When the optical thickness increases at fixed frequency, the total leakage rate of AM was seen to increase. Upon  varying the optical thickness we investigated separately the role of photonic spin   and  orbital momentum, the two essential constituents of electromagnetic AM \cite{cohen}, and  found both to co-exist.The torque on the source was also seen to increase with dipole density but hardly with optical thickness.
In this work we study the  frequency dependence of the dipole scatterers. Especially for large detuning from their resonance, the scattering from one dipole becomes weak so that to keep a reasonable optical thicknesses we  require more dipoles, typically thousands, that takes more CPU time and memory.

  \section{Leakage of Angular Momentum}

For a monochromatic  electric dipole source with electric dipole moment $\mathbf{d}$ at frequency $\omega = kc_0$, positioned at $\mathbf{r}=0$, the radiated electric field at position $\mathbf{r}$ is given by
\begin{equation}\label{E}
    \mathbf{E}(\mathbf{r},\omega) = - 4 \pi k^2 \mathbf{G}(\mathbf{r}, 0, \omega+i0)\cdot \mathbf{d}(\omega)
\end{equation}
with ${{G}}_{kk'} (\mathbf{r},\mathbf{r}', \omega)$  the vector Green's function associated with the Helmholtz equation for the electric field. This Green's function contains full information about the environment. The slightly positive imaginary part of the frequency $\omega+ i0$ guarantees outward propagation of the light.
The  power $P$ (radiated energy per second) radiated by the electric dipole is equal to its dissipation rate  $ \mathrm{Re}(\mathbf{J}^*\cdot \mathbf{E})/2$ \cite{Jackson}. Since $\mathbf{J}=  -i\omega \mathbf{d}$  we find, after averaging over the  orientation of the dipole source,

\begin{eqnarray}\label{PS}
P = -\frac{2\pi}{3} k^3c_0 |{\bf d}|^2 \mathrm{Im} \,\mathrm{ Tr} \, \mathbf{G}(0,0)
\end{eqnarray}
We recognize $\rho(k) \sim -\mathrm{Im} \,\mathrm{ Tr} \, \mathbf{G}(0,0)$ as the LDOS at the source position. The balance equation for the angular momentum can be written as \cite{OL},
\begin{eqnarray}\label{leak}
 \frac{d}{dt} {J}_{i, \mathrm{mec}} &=& {M}_i  \nonumber \\
 &=& \frac{R^3}{8\pi}  \epsilon_{ijk} \mathrm{Re }\int_{4\pi} d\hat{{\bf r} } \, \hat{r}_l \hat{r}_j(E_l^*E_k + B_l^*B_k)(R\hat{{\bf r} } ) \nonumber \\
\end{eqnarray}
with ${\mathbf{J}}_{\mathrm{mec}}$ the mechanical AM of the matter and with implicit summation over repeated indices. This formula expresses that the torque ${\bf M}$ exerted on the matter is radiated away as AM to infinity ($r>R)$, thereby assuming a source that has been constant during a time longer than $R/c_0$. In this picture  the radiative AM inside the environment enclosed by the sphere of radius $R$ is constant in time, and AM leaks to infinity somewhere ar $r(t) \sim c_0t > R$. The cycle-averaged torque acts on both source and its environment, ${\bf M} = {\bf M}_S + {\bf M}_E$. The latter is
\begin{equation}\label{ME}
{\bf M}_E = \frac{1}{2} \mathrm{Re} \, \int d^3{\bf r} \, \left[ {\bf P}^* \times  {\bf E} + P_m^* ({\bf r} \times  {\bf \nabla}) E_m \right]
\end{equation}
This torque vanishes for a rotationally-invariant environment around the source but not when it this symmetry is broken by structural heterogeneity, as will be discussed here. The torque on a source with an electric dipole moment $\mathbf{d}(\omega)$ is given by \cite{Jackson},
 \begin{equation}
{\bf M}_S =  \frac{1}{2} \mathrm{Re} \, (\mathbf{d}^* \times  {\bf E}  )
\end{equation}
With the electric field given by Eq.~(\ref{E}), an expression similar to Eq.~(\ref{PS}) can be obtained,
\begin{eqnarray}\label{MSS}
{M}_{S, i} = -\frac{2\pi}{3} k^2 |{\bf d}|^2 \mathrm{Re}\, \epsilon_{ijk} { G}_{kj}(0,0)
\end{eqnarray}
This torque vanishes for an (on-average) spherical environment with isotropic optical response \cite{LL}.

Alternatively, the leak of AM given by the righthand side of Eq.~(\ref{leak}) can be split up into parts associated with photonic spin  and orbital momentum \cite{cohen}
\begin{equation}\label{LS}
{\bf M} = \frac{R^2}{8\pi k } \mathrm{Im}   \int_{r=R} d^2{\bf \hat{r}} \, \left[ {\bf E}^* \times  {\bf E} + E_m^* ({\bf r} \times  {\bf \nabla}) E_m \right]
\end{equation}
Especially the existence of orbital AM expressed by the second term  is interesting since polarized radiation by a source subject to a magnetic field may be more intuitive to accept in view of the Faraday effect.

\section{Environment of $N$ Dipoles with Zeeman shift}
The Helmholtz Green function for light scattering from $N$ electric dipoles can be found in literature \cite{Glit}. Because of the pointlike nature of the dipoles it reduces to a $3N \times 3N$ complex-symmetric non-hermitian matrix. The magnetic response of the dipoles - due to the Zeeman splitting of their internal resonance - can be extracted in linear order so that the AM linear in the external field can be calculated numerically given the $N$ positions of the dipoles. The polarizability of a single dipole is given by,

\begin{eqnarray}\label{alpha}
{\bf \alpha} (\omega, \mathbf{B}_0) = \alpha(0) \frac{\omega_0^2}{\omega_0^2 - (\omega+ \omega_c i{\bf \epsilon\cdot \hat{B}}_0)^2 - i\gamma \omega}
\end{eqnarray}
in terms of the radiative damping rate $\gamma$, the resonant frequency $\omega_0$ and the cyclotron frequency $\omega_c = eB_0/2mc_0$. The second rank operator $i{\bf \epsilon\cdot \hat{B}}_0$ has the $3$ eigenvalues $0, \pm 1$ corresponding to $3$ Zeeman levels that make the environment linearly magneto-birefringent. The external magnetic field $\mathbf{B}_0$ is assumed homogeneous across the environment, but this can easily be altered in future work, e.g. to describe an environment surrounding a magnetic dipole . The detuning parameter is defined as $\delta \equiv (\omega -\omega_0)/\gamma $.  It is also useful to introduce the dimensionless material parameter $\mu = (12 \pi/\alpha(0) k^3) \times \omega_c/\omega_0$ that quantifies the magnetic birefringence induced by one dipole
and typically small (for the $1S$-$2P$ transition in atomic hydrogen one can estimate  $\mu \sim 7\cdot 10^{-5}$/Gauss ). The mean free path $\ell$ is another important length scale and follows from  $1/\ell = nk \mathrm{Im }\alpha(\omega)$ if we neglect recurrent scattering, with $n = 3N/4\pi a^3$ the density of dipoles. Without magnetic field the polarizability can be written as
$\alpha = -({3\pi }/{k^3})/(\delta + i/2)$.

\begin{figure}[t]
\includegraphics[width=8cm, angle=0]{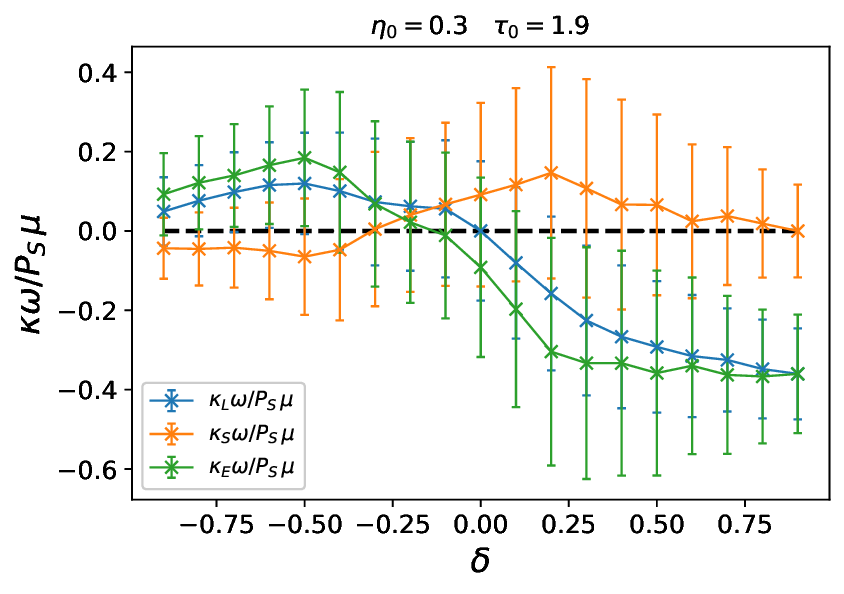}
\caption{Total normalized leakage rate of  angular momentum (blue) as a function of detuning $\delta= (\omega-\omega_0)/\gamma$ from the dipole resonance,  separated into torque on the source  (orange) and torque on the environment (green). The optical depth is $\tau=1.9$ and dimensionless dipole density is $\eta = 0.3$. }
\label{ESL}
\end{figure}

\begin{figure}
\includegraphics[width=8cm, angle=0]{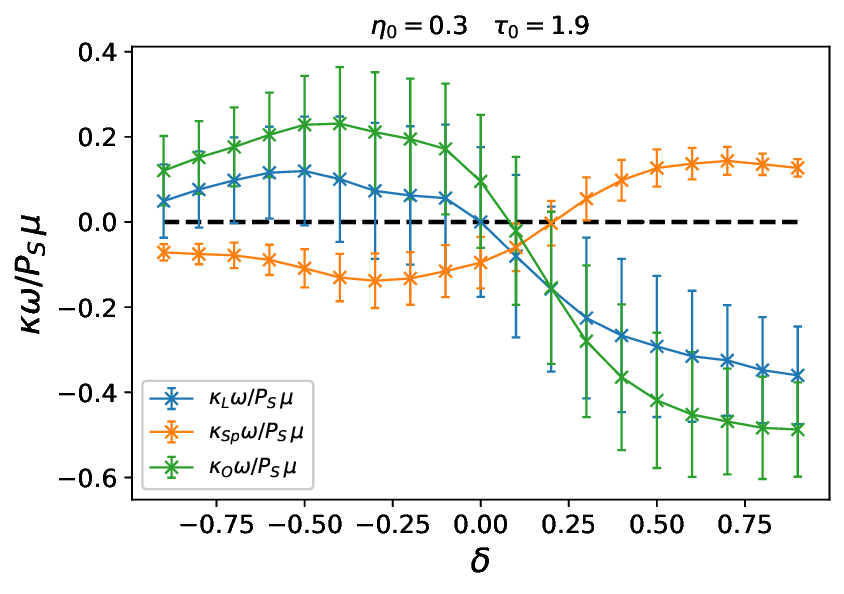}
\caption{As in previous figure with same fixed parameters $\tau=1.9$ and  $\eta = 0.3$,  but this time the total leakage has been split up into into leakage of orbital angular momentum  (green) and leakage of spin (orange). }
\label{SpOL}
\end{figure}

The leak of AM is calculated by performing the surface integral in Eq.~(\ref{leak}) numerically for different realizations in which dipole positions are averaged over a sphere with given radius $a $ at homogeneous average density throughout the sphere. For more details we refer to Ref.~\cite{PRBAM}. It was explicitly checked that the surface integral did not depend on the choice of $R>a $ as required by conservation of AM. Once verified it is convenient to evaluate
 Eq.~(\ref{leak}) in the far field  $R \gg a$ where the fields simplify. Our code was also tested on flux conservation and obeys the Optical Theorem.

\section{Numerical Results}
In this paper we focuss on numerical results obtained for different detunings $\delta$  and relatively large optical depths $\tau \equiv a/\ell$ in the hope to see major trends that can be extrapolated to even  larger detunings. This regime becomes  rapidly challenging since $\tau \sim \frac{9}{4} N/ (ka \times \delta)^2  $ so that for $\delta \gg 1$ and $\tau  \gg 1$ we need large $N$. Typically,  for  $\delta =2$, the best we have done sofar, and $\tau= 5$  we already need $N=2000$ in a sphere of 13 inverse wave numbers in radius  ($ka= 13$). These numbers imply a value for the number of dipoles per optical volume $\eta  \equiv 4 \pi n/k^3 = 3N/(ka)^3 \sim 2.5 $, i.e. the dipoles are largely located in each others near field. Nevertheless for a detuning $\delta =2$, dipoles still scatter more or less independently because
$ k\ell \sim  2\delta^2/ \eta= 3 > 1$ \cite{pr} but this is no longer true for $\eta=6$. This implies that completely unknown effects such weak localization may affect the radiative transfer of AM.

After an ideal average over all $N$ dipole positions, the magnetic field is the only orientation left in the problem and we expect that $\mathbf{M}= \kappa \mathbf{\hat{B}}_0$ with $\kappa$ a real-valued scalar to be calculated that can have both signs. Following earlier work \cite{pinwheel, OL, PRBAM} we normalize the leakage of AM by the radiated amount of energy  and introduce the dimensionless AM $\kappa \omega/P$, with $P$ the radiated amount of energy per second. This number is linear in the material parameter $\mu$ introduced earlier and can directly be related to the Hall angle of the Poynting vector in the far field of the sphere. Alternatively, the number quantifies the amount of leaked angular momentum expressed in $\hbar$, normalized per emitted photon.

In Figures~\ref{ESL} and \ref{SpOL} we show the normalized AM leakage for an optically thin sphere as a function of detuning. The bars in all figures denote the typical support of the full PDF when calculating the torque for $1000$ different realizations of the dipole positions. Except for the  spin leakage rate, they are large and all AM  related to source and orbital momentum are genuine mesoscopic parameters. The optical depth $\tau= 1.9 $ and average density $\eta= 0.3$ are kept constant which means that both the number $N$ of dipoles and the radius $a$ of the sphere change as $\delta$ is varied. It is seen that the dimensionless AM depends significantly on detuning and changes sign near the resonance at $\delta= 0$. In this weakly scattering regime single scattering is still dominant. For a thin layer in the far field of  the source we can derive a profile  $\kappa \omega/P \sim \eta\,  \mathrm{ Im} \alpha^2 \sim -\eta \delta/(\delta^2 + 1/4)^2$ independent of distance. This corresponds more or less to the observed profile of total AM in Figures~\ref{ESL} and \ref{SpOL} that are nevertheless affected by higher order scattering events. Both figures also show how total AM leakage splits up into either spin $+$ orbital AM (Figure~\ref{ESL}) or torque on source  $+$ torque on environment (Figure \ref{SpOL}). All are of same order of magnitude but are not always of same sign. In particular for this set of parameters both decompositions have opposite sign.

\begin{figure}[t]
\includegraphics[width=8cm, angle=0]{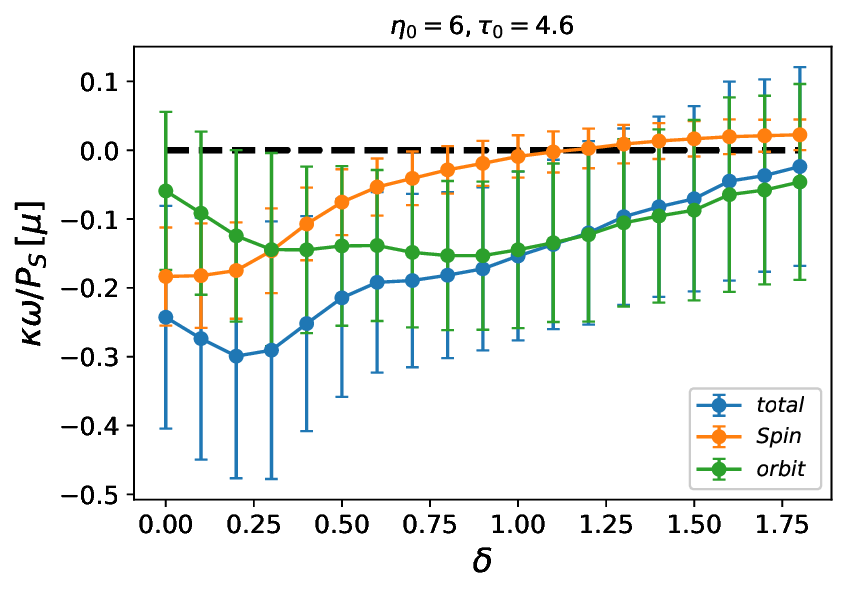}
\caption{Total leakage of optical angular momentum (blue) as a function of detuning. The orange and green curves represent spin and orbaital momentum. The optical depth and average density are kept constant ($\tau=4.6$ and  $\eta = 6$). The calculations have  been done only for $\delta > 0$, i.e. blue-shifted from the resonance.}
\label{SpOL2}
\end{figure}

\begin{figure}[t]
\includegraphics[width=8cm, angle=0]{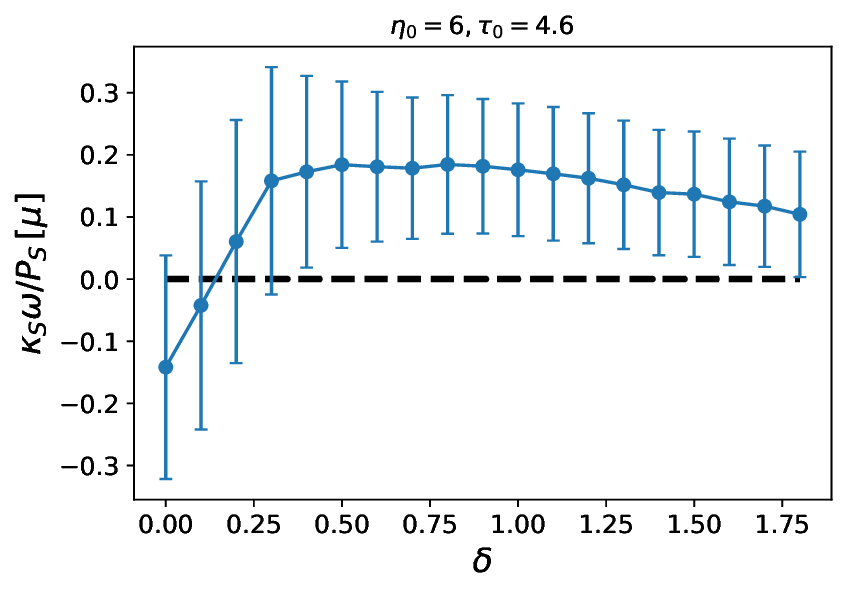}
\caption{As in Figure \ref{SpOL2}, for clarity only the torque on the source has been shown. Bars denote the support of the full probability distribution (PDF) over 1000 realizations.  }
\label{Source}
\end{figure}

The picture changes significantly in the multiple scattering regime. Figures~\ref{SpOL2} and \ref{Source} show the same normalized leakage of AM for optical depth $\tau= 4.6$ and dimensionless density as large as $\eta=6$. Going to smaller values of $\eta$ would require a too large value for $N$. The total leakage is now negative for all (positive) detunings and spin leakage and orbital leakage have same sign. Except near resonance, it is dominated by leaks in orbital AM. From Figure~\ref{Source} we see that the torque on the source is mainly positive but changes sign with detuning near resonance. This implies again that for most detunings source and environment are subject to opposite torques. A normalized torque on the source around $0.1-0.2$ is not much different than what was found for $\eta=0.3$ \cite{PRBAM}. For low densities this torque increased with $\eta$ but seems to saturate for $\eta > 0.3$.   The numbers for total leakage rate ($-0.2 \pm 0.1$) are almost one order of magnitude less than what we found for $\eta= 0.03$ and $\tau\approx 2$  in Ref.~\cite{PRBAM} and one may speculate about the possibility of some process related to ``weak localization"  that reduces the transfer of angular momentum.

To get an order of magnitude for this effect we consider an homogenous gas with atoms of mass $Z \times m_H$ in a sphere of size $a$. For an ideal gas at room temperature the density is roughly $n_0=40$ mol/$\mathrm{m}^3$. If we assume that all leaked angular momentum after a time interval $\Delta t$ is transferred homogeneously to mechanical momentum of the sphere, we can estimate that its angular velocity is
\begin{eqnarray}\label{Omega}
    \Omega [\mathrm{rad}/\mathrm{s}^{-1} ] &= &5.7 \frac{\kappa \omega}{P \mu } \times\frac{ \mu}{Z} \times
   \frac{ \left( \frac{P}{10\, \mathrm{W}}\right) \left( \frac{\lambda}{500\,  \mathrm{nm} }\right) \left( \frac{\Delta t}{\mathrm{days}}\right)}{ \left( \frac{a}{1 \, \mathrm{cm}}\right)^5 \left( \frac{n}{n_0}\right)} \nonumber
\end{eqnarray}
For $\kappa \omega/P\mu =0.3 $, $\mu/Z \sim 10^{-5}$ the angular rotation is typically of the order of $1$ mrad/s after 100 days.

\section{Conclusion}
In this work we have reported an exact  numerical study of the radiation  of  electromagnetic angular momentum by a light source imbedded in a  disordered and magneto-active environment described by resonant electric dipoles with Zeeman splitting that scatter light elastically. The angular momentum is directed along the magnetic field and its transfer is directed radially outward from the source.  It is in general composed of both spin transport and transport of orbital angular momentum. The first implies polarization of radiated light, the second is related to an energy flux circulating around the object and the magnetic field.  Leakage of angular momentum  has been quantified by a dimensionless parameter that is essentially the ratio of angular momentum leakage rate (with physical unit Joule) and the energy of the source emitted during one optical cycle.
The number can be seen to be equal to the angular momentum, expressed in $\hbar$, transferred per emitted photon to source and environment. It is proportional to the product of a pure material parameter $\mu$ associated with the magneto-scattering of the dipoles and the numbers that can be found in the figures that result from scattering.  By conservation of angular momentum this transfer gives rise to torques on both the emitting source and the scattering environment. All parameters are seen to be of the same order, can have mutually opposite signs and depend on the detuning from the resonance.

The regime of Thompson scattering is interesting for astrophysical application and is a major scattering mechanism in our Sun. It corresponds to large detunings, where the phase shift  between incident and scattered field becomes negligible.  Out-of phase response seems essential for the leakage to exist, even for large optical depths. The amount of multiple scattering, quantified by optical depth,  certainly affects radiative transfer of angular momentum but it is difficult at this point to deduce general trends. Our simulations are clearly in need of a radiative transport theory for  angular momentum in magnetic fields that, to our knowledge, does not exist. From our simulations we suspect that different parts of the environment undergo different torques. We also expect that as the optical thickness of the environment increases, the precise nature of the source becomes of less importance, quite opposite to what was found for a homogeneous environment.

Indeed, this picture may possibly apply to stellar atmospheres, globally exposed to the magnetic dipole fields of their nuclei. Although all ingredients are present for leakage of angular momentum to exist, lots of extra complications, such as broadband radiation, Doppler broadening, etc make it difficult for quantitative predictions.

  \end{document}